\documentclass[twocolumn,notitlepage,aps,pra,superscriptaddress,amsmath,amssym,floatfix,longbibliography]{revtex4-1}
\usepackage{subcaption}
\usepackage{graphicx}
\usepackage{bm}
\usepackage{color}
\usepackage{amssymb}
\usepackage{enumerate}
\usepackage{comment}
\usepackage{amsmath,amssymb}
\usepackage{tikz-cd}

\def\la{\langle}
\def\ra{\rangle}

\def\be{\begin{equation}}
\def\ee{\end{equation}}

\begin{document}

\newcommand{\bigjprob}{{\mathcal{P}}}
\newcommand{\bigprob}{_{\bm{q}_F}{\mathcal{P}}_{\bm{q}_I}}
\newcommand{\cum}[1]{\llangle #1 \rrangle}       					
\newcommand{\op}[1]{\hat{\bm #1}}                					
\newcommand{\vop}[1]{\vec{\bm #1}}
\newcommand{\opt}[1]{\hat{\tilde{\bm #1}}}
\newcommand{\vopt}[1]{\vec{\tilde{\bm #1}}}
\newcommand{\td}[1]{\tilde{ #1}}
\newcommand{\mean}[1]{\la#1\ra}                  					
\newcommand{\cmean}[2]{ { }_{#1}\mean{#2}}       				
\newcommand{\pssmean}[1]{ { }_{\bm{q}_F}\mean{#1}_{\bm{q}_I}}
\newcommand{\ket}[1]{\vert#1\ra}                 					
\newcommand{\bra}[1]{\la#1\vert}                 					
\newcommand{\ipr}[2]{\left\la#1\mid#2\right\ra}            				
\newcommand{\opr}[2]{\ket{#1}\bra{#2}}           					
\newcommand{\pr}[1]{\opr{#1}{#1}}                					
\newcommand{\Tr}[1]{\text{Tr}(#1)}               					
\newcommand{\Trd}[1]{\text{Tr}_d(#1)}            					
\newcommand{\Trs}[1]{\text{Tr}_s(#1)}            					
\newcommand{\intd}[1]{\int \! \mathrm{d}#1 \,}
\newcommand{\dd}{\mathrm{d}}
\newcommand{\fullint}{\iint \! \mathcal{D}\mathcal{D} \,}
\newcommand{\drv}[1]{\frac{\delta}{\delta #1}}
\newcommand{\partl}[3]{ \frac{\partial^{#3}#1}{ \partial #2^{#3}} }		
\newcommand{\smpartl}[4]{ \left( \frac{\partial^{#3} #1}{ \partial #2^{#3}} \right)_{#4}}
\newcommand{\smpartlmix}[4]{\left( \frac{\partial^2 #1}{\partial #2 \partial #3 } \right)_{#4}}
\newcommand{\limit}[2]{\underset{#1 \rightarrow #2}{\text{lim}} \;}
\newcommand{\funcd}[2]{\frac{\delta #1}{\delta #2}}
\newcommand{\funcdiva}[3]{\frac{\delta #1[#2]}{\delta #2 (#3)}}
\newcommand{\funcdivb}[4]{\frac{\delta #1 (#2(#3))}{\delta #2 (#4)}}
\newcommand{\funcdivc}[3]{\frac{\delta #1}{\delta #2(#3)}}
\definecolor{dgreen}{RGB}{30,130,30}

\newcommand{\Dave}[1]{{\color{magenta} #1}}

\newcommand{\ams}[1]{{\color{teal} #1}}
\newcommand{\Andrew}[1]{{\color{red} #1}}

\title{Theory of direct measurement of the quantum pseudo-distribution via its characteristic function}

\author{Andrew N. Jordan}
\affiliation{Institute for Quantum Studies, Chapman University, Orange, CA 92866, USA}
\affiliation{Department of Physics and Astronomy, University of Rochester, Rochester, NY 14627, USA}
\email{jordan@chapman.edu}
\author{David  R. M. Arvidsson-Shukur}
\affiliation{Hitachi Cambridge Laboratory, J. J. Thomson Avenue, Cambridge CB3 0HE, United Kingdom}
\affiliation{Institute for Quantum Studies, Chapman University, Orange, CA 92866, USA}
\author{Aephraim M. Steinberg}
\affiliation{Department of Physics, University of Toronto, 60 St. George Street, Toronto ON, M5S 1A7, Canada}
\affiliation{Institute for Quantum Studies, Chapman University, Orange, CA 92866, USA}

\date{\today}

\begin{abstract}
We propose a method for directly measuring the quantum mechanical pseudo-distribution of observable properties via its characteristic function.  Vandermonde matrices of the eigenvalues play a central role in the theory. This proposal directly finds the pseudo-distribution using weak measurements of the generator of position moments (momentum translations).  While the pseudo-distribution can be extracted from the data in a theory-agnostic way, it is shown that under quantum-mechanical formalism, the predicted pseudo-distribution is identified with the Kirkwood-Dirac pseudo-distribution.  We discuss the construction of both the joint pseudo-distribution and a conditional pseudo-distribution, which is closely connected to weak-value physics. By permuting position and momentum measurements, we give a prescription to directly probe the canonical commutation relation and verify it for any quantum state.
This work establishes the theory of a characteristic function approach to pseudo-distributions, as well as providing a constructive approach to measuring them directly.
\end{abstract}

\maketitle
\section{Introduction} 

The non-commutation of certain observables in quantum physics indicates that the joint assignment of their properties cannot be made unambiguously --- a measurement of one observable is guaranteed to modify the value of the other \cite{jordan2024quantum}. 
Consequently, one generally cannot describe a quantum state as a joint probability distribution with respect to several observables. One can, however represent a quantum state by a pseudo-distribution with respect to non-commuting observables. A pseudo-distribution is similar to a joint probability distribution: it normalises to $1$ and it produces the correct Born-rule probabilities.  
However, a pseudo-distribution may have negative or non-real entries.  

There are many pseudo-distributions. Examples include the Wigner function \cite{wigner1932quantum}, the Kirkwood-Dirac  \cite{kirkwood1933quantum,terletsky1937limiting,dirac1945analogy} distributions, the Husimi ``Q'' function \cite{husimi1940some}, and the Glauber-Sudarshan ``P'' function \cite{glauber1963coherent,sudarshan1963equivalence}.  A pseudo-distribution, together with a corresponding description of transformations and measurements, forms a representation of quantum experiments. 
Such pseudo-probability descriptions have been found useful in a wide variety of contexts, including ``direct'' measurements of the wavefunction \cite{lundeen2011direct} and density matrix \cite{thekkadath2016direct}, quantum metrology \cite{hofmann2011uncertainty,arvidsson2020quantum,lupugladstein2022metrology}, weak value amplification \cite{aharonov1988result,hosten2008observation,dixon2009ultrasensitive, dressel2015weak,dressel2014colloquium,jordan2024quantum}, quantum thermodynamics \cite{yunger2017jarzynski,levy2020quasiprobability}, quantum optics in phase space  \cite{shalm2009squeezing,schleich2015quantum}, 
and foundational issues in quantum mechanics \cite{mir2007whichpath,williams2008weak,bednorz2012nonclassical,pusey2014anomalous,kunjwal2019anomalous}.  For a detailed recent review of the Kirkwood-Dirac distribution and its applications, see Ref.~\cite{arvidsson2020quantum}.

While pseudo-probabilistic representations of quantum mechanics have been of tremendous use, several puzzles remain.
First, why are there many of them?  When analyzing quantum experiments, which pseudo-distribution should be applied? These questions are particularly interesting with respect to a well known result connecting genuine non-classical phenomena (in the form of generalised contextuality \cite{spekkens2005contextuality}) to negativity in quasi-probabilistic representations. If an experiment's preparation, transformation and measurement cannot be described by any positive quasi-probabilistic representation, then that experiment is genuinely non-classical \cite{spekkens2008negativity}. In general, this result is hard to apply; there are an infinite number of pseudo-distribution to check.  However, might some experiments admit a theory-independent way to reconstruct a unique pseudo-distribution from measured data, without a priori assumptions of its form?

We consider an experiment with three consecutive events in time: prepare, measure, and re-measure. (1) A quantum system is prepared in an initial state $\psi$. (2) 
A non-invasive generalized measurement (e.g. a ``weak measurement''\cite{aharonov1988result}) measures the 
expectation value $\langle \hat{A}^n \rangle$
of some power $n$ of an observable $\hat{A}$. We label the physical (eigen)values (assumed non-degenerate) of the operator by $a_i$. 
(3) The system is subject to a projective fine-grained measurement in some basis $\{ \ket{b_j}\}$.
Further, we allow the experimentalist to post-select the data conditioned on the outcomes from the final measurement, thus applying a final boundary condition. In other words, the experimentalist can ask: What is the expectation value of observable $\hat{A}^n$ conditioned on preparation in state $\ket{\psi}$ and a final outcome corresponding to $\ket{\phi}=\ket{b_{j'}}$?

If the experiment were classical, one would describe the pre- and post-selected system with some conditional probability distribution $P_{i|j}(\psi)$. For all $n$, the conditional expectation values would be given by $\sum_{i} P_{i|j}(\psi) a_i^n$. In quantum mechanics, instead, the pre- and post-selected system can be described by a conditional pseudo-distribution $Q_{i|j}(\psi)$. But what specific form does $Q_{i|j}(\psi)$ take? And how can we infer the generally complex-valued $Q_{i|j}(\psi)$ from experiments?

In this Article, we address both these questions. We show that if one requires that the form of the classical conditional expectation value apply also to the quantum scenario, then $Q_{i|j}(\psi)$ takes a unique form. In other words, if the conditional expectation value of $\hat{A}^n$ is to be given by $\sum_{i} Q_{i|j}(\psi) a_i^n$, then we predict that the experimental data forces $Q_{i|j}(\psi)$ to be a Kirkwood-Dirac pseudo-distribution (defined below). Furthermore, we detail how measurements of the powers of an operator enable the determination of the underlying quasi-probabilistic characteristic function.  This enables the direct inference of the the underlying pseudo-distribution. We then extend our analysis from conditioned expectation values to general correlation functions, and give a prescription to uniquely determine more general pseudo-distributions.
Our efficient method is in contrast to previous methods to measure pseudo-distributions, which relied on scanning weak measurements over all eigenprojectors of $\hat{A}$ and sharp measurements in the $\{ \ket{b_j} \}$ basis \cite{arvidsson2024properties}.

The remainder of this article is organized as follows.  In Sec.~\ref{Background} we review some important background information about properties of pseudo-distributions in quantum physics, focusing on the Kirkwood-Dirac distribution.  In Sec.~\ref{finite}, we explore finite dimensional Hilbert spaces, and make an analysis of constructing pseudo-distributions using data from a finite set of moment measurements of a given observable.  Connections to weak values are made. In subsection \ref{qubit-sec}, a qubit example is given to illustrate the theory, and we show how the conditional pseudo-distribution is related to weak values of eigenstate projectors. In subsection \ref{two-observable}, we extend the analysis to measuring moments of two (non-commuting) observables and finding the joint pseudo-distribution of the two outcomes. Generalizations are made to an arbitrary number of measurements in subsection \ref{multiple}.  In Sec.~\ref{continuous}, we consider continuous variable systems and develop a generalization of our discrete variable method.  We show how a characteristic function can be defined and that the pseudo-distribution can be calculated via its Fourier transform.  In Sec.~\ref{experiment} an experiment is proposed that directly measures the generating function of moments of position, with a subsequent measurement of momentum.  Methods for finding both conditional and joint pseudo-distributions are given.  We also give a prescription for validating the canonical commutation relation for any state.

\section{Background}
\label{Background}

Before we present our analysis, we outline some theoretical background material. For now, we work on a finite Hilbert space $\mathcal{H}$ of dimensionality $d$. Extensions to continuous-variable systems are straightforward.  Consider two non-degenerate observables represented by the Hermitian operators $\hat{A}= \sum_i a_i \ket{a_i}\bra{a_i}$ and $\hat{B}= \sum_j b_j \ket{b_j}\bra{b_j}$. Here, $a_i$ and $b_j$ denote eigenvalues and $\ket{a_i}$ and $\ket{b_j}$ their corresponding eigenvectors. In what follows, we assume for simplicity that $\hat{A}$ and $\hat{B}$ are fully incompatible in the sense that $\la a_i | b_j \ra \neq 0$ for all $i,j$. Violations of this assumption can be handled by infinitesimal tilts of the operators.

The observable bases of $\hat{A}$ and $\hat{B}$ provide alternative ways of representing a quantum state $\hat{\rho}$. For example, one can define two related bases, known as the \textit{primary frame} and the \textit{dual frame}, for operators on $\mathcal{H}$:
\begin{equation}
    \hat{F}_{i,j} = \ket{a_i} \la a_i | b_j \ra \bra{b_j} , \; \text{and} \;  \hat{G}_{i,j} = \frac{ \ket{a_i}  \bra{b_j}}{\la b_j | a_i \ra} . 
\end{equation}
We note that $\Tr{\hat{F}_{i,j} \hat{G}^{\dagger}_{i',j'} } = \delta_{i,i'} \delta_{j,j'}$. 
The Kirkwood-Dirac pseudo-probability representations for a state $\hat{\rho}$ and an observable $\hat{X}$ are then given by
\begin{equation}
\label{KD}
    K_{i,j}(\hat{\rho}) = \Tr{\hat{F}_{i,j}^{\dagger} \hat{\rho}} = \la b_j | a_i \ra \la a_i | \hat{\rho} | b_j \ra  ,
\end{equation}
and
\begin{equation}
    T_{i,j}(\hat{X}) = \Tr{\hat{G}_{i,j}^{\dagger} \hat{X}} = \frac{ \la a_i | \hat{X} | b_j \ra }{\la a_i | b_j \ra} ,
\end{equation}
respectively. $K_{i,j}(\hat{\rho})$ and $T_{i,j}(\hat{X})$ are informationally complete. 
For example, if one knows $K_{i,j}(\hat{\rho})$, one can reconstruct $\hat{\rho}$:
\begin{equation}
    \hat{\rho} = \sum_{i,j} K_{i,j}(\hat{\rho}) \hat{G}_{i,j} . \label{reconstruct}
\end{equation}
$K_{i,j}(\hat{\rho})$ is normalized and produces the correct Born-rule probabilities:
\begin{equation}
     \sum_{j} K_{i,j}(\hat{\rho}) = \la a_i | \hat{\rho} | a_i \ra, \ \sum_{i} K_{i,j}(\hat{\rho}) = \la b_j | \hat{\rho} | b_j \ra, 
\end{equation}
and the joint sum giving unity.
Moreover, the expectation value of $\hat{X}$ with respect to $\hat{\rho}$ is given by a quasi-probabilistic weighted average:
\begin{equation}
    \Tr{\hat{X} \hat{\rho}} = \sum_{i,j} T_{i,j}^{\dagger}(\hat{X}) K_{i,j}(\hat{\rho}) .
\end{equation}
For some states, $K_{i,j}(\hat{\rho}) \in [0,1]$ for all $i,j$. Then, $K_{i,j}(\hat{\rho})$  equals a standard joint probability distribution. However, in general, $K_{i,j}(\hat{\rho})$ takes values in the complex unit disc. Non-real and negative values in $K_{i,j}(\hat{\rho})$ have been connected to a range of non-classical phenomena \cite{arvidsson2024properties}. 

The Kirkwood-Dirac distribution $K_{i,j}$ satisfies a quasi-probabilistic version of Bayes' theorem (different to the version exhibited by SIC-POVMs \cite{Medendorp2011POVM}). One can condition $K_{i,j}$ on a specific eigenvalue of the $\hat{B}$ observable. The conditional distribution is
\begin{equation}
\label{Eq:CondKD}
    K_{i|j}(\hat{\rho}) = K_{i,j}(\hat{\rho}) / \la b_j| \hat{\rho} | b_j \ra.
\end{equation}
If $\hat{\rho} = \ket{\psi}\bra{\psi}$ is pure, then
\begin{equation}
\label{CondKD}
    K_{i|j}(\psi) = \frac{\la b_j | a_i \ra \la a_i | \psi \ra }{\la b_j | \psi \ra} .
\end{equation}
$K_{i|j}(\psi)$ is a \textit{weak value} of the projector observable $|a_i\ra \la a_i|$ \cite{aharonov1988result,dressel2014colloquium,jordan2024quantum}. Conceptually, $K_{i|j}(\psi)$ represents the conditional expectation value of $|a_i\ra \la a_i|$ with respect to an initial state $\ket{\psi}$ that is post-selected to be in the state $\ket{b_j}$. 
The weak value of a general observable $\hat{C}$ is 
\begin{equation}
    \la \hat{C} \ra^{\psi}_{b_j} = \frac{\la b_j | \hat{C} | \psi \ra }{\la b_j | \psi \ra} . \label{wv}
\end{equation}
Weak values can be measured as conditioned averages in the weak measurement limit \cite{aharonov1988result,dressel2014colloquium,jordan2024quantum}. From Eq.~\eqref{Eq:CondKD} it is clear that if one measured  $K_{i|j}(\hat{\rho})$ for all $i,j$ and the marginal probabilities $\la b_j | \rho | b_j \ra$, then one could reconstruct $K_{i,j}(\hat{\rho})$, and thus $\hat{\rho}$. However, this requires the estimation of the $d \times d$ combinations of all $i$ and $j$. Below, we will outline a methodology for measuring $d-1$ post-selected expectation values of the powers of $\hat{A}$. 
This methodology uniquely determines all the entries of any reasonable choice of $Q_{i,j}$ to represent the data. Moreover, we find that this unique choice of $Q_{i,j}$ is the Kirkwood-Dirac distribution $K_{i,j}$. 
 
\section{Pseudo-Distribution Construction method}
\label{finite}
We consider many repetitions of the following three-step experiment. First, we prepare a quantum system in an initial state $|\psi\ra$. Second, we conduct a weak (infinitesimally perturbing) measurement of the system's expectation value $\la \hat{A}^n \ra $, where we vary $n$ from $0$ to $d-1$, where $d$ is the dimensionality of the Hilbert space. Third, we conduct a strong measurement of the observable $\hat{B}$ such that the quantum system is projected onto a vector in the $\{ \ket{b_j} \}$ basis. Quantum theory tells us that the observed expectation value of $\hat{A}^n$ with respect to an initial state $\ket{\psi}$ and a final state $\ket{b_j}$ is given by Eq.~(\ref{wv}) with ${\hat C} = {\hat A}^n$. 
Classically, a conditional expectation value of $\hat{A}^n$ is $\la {\hat A}^n\ra^{\psi}_{b_j} = \sum_{i=1}^d (a_i)^n P_{i|j}$ for all $n$. Here, $P_{i|j}$ is the classical system's conditional probability distribution.  Consequently, if we wish to describe the quantum-mechanical conditional expectation value of any moment of an operator [Eq.~\eqref{wv}] with a pseudo-statistical theory where $Q_{i|j}$ takes the role of $P_{i|j}$, then we have that
\begin{eqnarray}
\label{moments}
    \la {\hat A}^n\ra^{\psi}_{b_j} = \sum_{i=1}^d (a_i)^n Q_{i|j} = \frac{\la b_j |{\hat A}^n|\psi\ra }{\la b_j |\psi\ra } , \text{ for all } n,
\end{eqnarray}
Here, the first equality follows from the assumption that {\it any} conditional pseudo-distribution exists, while the second follows from quantum weak-value theory.
Next, we show that such a $Q_{i|j}$ exists that reproduces the data from our three-step experiment. Moreover, we show how to use our experiment to uniquely determine $Q_{i|j}$.


Equation~(\ref{moments}) can be represented as a matrix equation. Let us define a $d$-dimensional column vector ${\vec A}$ that we use the organize the conditioned data. The $k$th element of ${\vec A}$ is the (conditioned) average of $\hat{A}^{k-1}$ [Eq. \eqref{wv}]. That is, ${\vec A}_k = \la {\hat A}^{k-1}\ra^{\psi}_{b_j}$, where $k \in [1, \ldots, d]$.  Similarly, we define the column vector ${\vec Q}$ such that $\vec{Q}_k = Q_{k|j}$, where $k \in [1, \ldots, d]$ are the elements of the pseudo-distribution.   
We then have the matrix equation 
\begin{equation}
\label{matrixeq}
    {\vec A} = {\bf V} \cdot {\vec Q} .
\end{equation}
 Here, we have defined a real-valued Vandermonde matrix ${\bf V}$ \cite{macon1958inverses} consisting of powers of the eigenvalues $a_i$: 
\begin{equation}
  {\bf V}=   \begin{pmatrix} 1 & 1 & \ldots  & 1 \\
a_1 & a_2 & \ldots & a_{d} \\
a_1^2 & a_2^2 & \ldots  & a_{d}^2 \\
\vdots & \vdots &   & \vdots \\
a_1^{d-1} & a_2^{d-1} & \ldots & a_{d}^{d-1} 
\end{pmatrix} 
\end{equation}
 That is, we have
\be
\begin{pmatrix}
1 \\
\la {\hat A}\ra^{\psi}_{b_j} \\
\la {\hat A}^{2}\ra^{\psi}_{b_j} \\
\vdots \\
\la {\hat A}^{d-1}\ra^{\psi}_{b_j}
\end{pmatrix}
 =\begin{pmatrix} 1 & 1 & \ldots  & 1 \\
a_1 & a_2 & \ldots & a_{d} \\
a_1^2 & a_2^2 & \ldots  & a_{d}^2 \\
\vdots & \vdots &   & \vdots \\
a_1^{d-1} & a_2^{d-1} & \ldots & a_{d}^{d-1} 
\end{pmatrix} \begin{pmatrix} Q_{1|j} \\
Q_{2|j} \\ Q_{3|j} \\ \vdots \\ Q_{d|j}\end{pmatrix}  .
\ee

Given empirical data of the real and imaginary parts of  ${\vec A}$, we can then make a matrix inversion in order to find the empirical pseudo-distribution:
\begin{equation}
    {\vec Q} = {\bf V}^{-1} \vec{A} .
\end{equation}
${\bf V}$ is invertible if its Vandermonde determinate,
\be
V = \prod_{1 \le i < j \leq d} (a_j-a_i),
\ee
is non-zero \cite{macon1958inverses}, so we assume a non-degenerate operator.
The matrix elements of the inverse matrix ${\bf V}^{-1}$ are given by the coefficients of the power-series expansion of a characteristic polynomial, with efficient numerical inversion possible in a time that scales as $N^2$ \cite{press2007numerical}.
More precisely, the inverse matrix elements are coefficients in the power series expansion of $d$ Lagrange interpolating polynomials $L_n(x)$   $(1\le n \le d)$ with the property  that $L_n(a_m) = \delta_{n,m}$.
We note that while we could go to lower or higher operator order, this will give rise to an under- or over-determined set of equations; the least squares solution can still be obtained via a pseudo-inverse.

To demonstrate the uniqueness of $Q_{i|j}$ we revisit Eq. \eqref{moments}. By inserting the spectral decomposition of $\hat{A}$ on the right-hand side, we find that 
\begin{equation}
\sum_i (a_i)^n Q_{i|j} = \sum_i (a_i)^n \frac{\la b_j | a_i \ra \la a_i |\psi\ra }{\la b_j |\psi\ra } =  \sum_i (a_i)^n K_{i|j} ,
\end{equation}
for all $n$. Here, we used Eq.~(\ref{CondKD}) in the last equality. Consequently, the conditioned Kirkwood-Dirac pseudo-distribution $K_{i|j}$ satisfies the desired formulae. The uniqueness follows from Eq.~\eqref{matrixeq}.


\subsection{Qubit example} \label{qubit-sec}

Let us consider a qubit ($d=2$) example.  We take the $\hat{A}$ observable to be the Pauli-$z$ operator: ${\hat A} = {\hat \sigma}_z$. Then, given the empirical value of $\la {\hat A} \ra_j$, Eq. \eqref{matrixeq} is 
\be
\begin{pmatrix} 1 \\
\la {\hat \sigma}_z\ra^{\psi}_{b_j} \end{pmatrix}=\begin{pmatrix} 1 & 1 \\
1 & -1   
\end{pmatrix} \begin{pmatrix} Q_{1|j} \\
Q_{2|j} \end{pmatrix}.
\ee
By inverting the Vandermonde matrix we find the underlying conditioned pseudo-distribution:
\be
\begin{pmatrix} Q_{1|j} \\
Q_{2|j} \end{pmatrix} = \frac{1}{2}
\begin{pmatrix} 1+\la {\hat \sigma}_z\ra^{\psi}_{b_j}  \\
1-\la {\hat \sigma}_z\ra^{\psi}_{b_j} \end{pmatrix}.
\ee
We note that if the weak value of ${\hat \sigma}_z$ exceeds the eigenvalues, then the pseudo-distribution elements $Q_{i=1,2|j}$ must be negative or exceed 1. 
This makes a clear connection between weak values that are complex or exceed the eigenvalue range to non-classical pseudo-distributions that exceed the classical range of $[0,1]$, see also  Ref.~\cite{pusey2014anomalous, yunger2018quasiprobability}. 

Further insight can be made by noting that the projection operators of the basis states  $\ket{a_1}$ and $\ket{a_2}$ can be written as $\ket{a_i}\bra{a_i} =  ( {\bf 1} + (-1)^{i+1} \sigma_z )/2$, for $i =1,2$. Thus,  the pseudo-distribution produced by this method is equal to 
\be
Q_{i|j} = \frac{ \la b_j | ( \ket{a_i}\bra{a_i} )  |\psi\ra}{\la b_j |\psi\ra }, \text{ for } i=1,2.
\ee
As discussed above, we find that $Q_{i|j} = K_{i|j}$. In other words, the conditional pseudo-distribution coincides with the conditional Kirkwood-Dirac pseudo-distribution \cite{arvidsson2024properties}. Consequently, the requirement of reproducing the conditional expectation values [Eq. \eqref{moments}] using the standard statistical formulae imposes the Kirkwood-Dirac distribution as the unique pseudo-distribution.

\subsection{Two-observable measurements}
\label{two-observable}
In the analysis above, we focused on the measurement of a single observable $\hat A$ conditioned on an initialization in $\ket{\psi}$ and a final measurement in one of the $\{ \ket{b_j} \}$ basis' states. Thus, we showed how to uniquely  determine and measure the conditional Kirkwood-Dirac pseudo-distribution $K_{i|j}(\psi)$.  We can extend this analysis to account for unconditional measurements, such that the ``standard'' Kirkwood-Dirac pseudo-distribution [Eq. \eqref{KD}] appears.  Consider a correlation function of the $n$th moment of operator ${\hat A}$ and the $m$th moment of operator ${\hat B}$.  The set of correlation functions is given by
\be
\label{cnm}
C_{n,m} = \la \psi| {\hat A}^n {\hat B}^m | \psi\ra.
\ee
Here we consider the system in a pure state $\psi$, but extensions to mixed states are straightforward.  We note that there are several ways to measure the correlation functions. One could use a three-step experiment: (1) Initialize a quantum system in $\ket{\psi}$. (2) Weakly measure the observable $\hat{B}^n$. (3) Projectively measure the observable $\hat{A}^m$.  We note that the operator ordering is switched from the previous section for aesthetic reasons. Alternatively, one can reconstruct the correlator from separate measurements of the Hermitian observables $i({\hat A}^n {\hat B}^m-  {\hat B}^m {\hat A}^n)$ and ${\hat A}^n {\hat B}^m + {\hat B}^m {\hat A}^n$ \cite{buscemi2013direct}.

Classically, the correlation functions would be calculated from an underlying joint probability distribution $P_{i,j}$ such that $C_{n,m} = \sum_{i,j} a_i^n b_j^m P_{i,j}$. Again, we transfer this statistical structure to the quantum setting and invoke a pseudo-distribution $Q_{i,j}(\psi)$. 
Thus, the correlation function is given by
\be
C_{n,m}  = \sum_{i,j} a_i^n b_j^m Q_{i,j}(\psi). \label{cnmsum}
\ee
Moreover, from quantum mechanics we have that the correlation functions can be defined via the {\it characteristic function} $Z(\lambda, \chi)$:
\be
C_{n,m} = (-i)^{n+m}\partial_\lambda^n \partial_\chi^m Z(\lambda, \chi)|_{\lambda = \chi =0},
\ee
where 
\be
\label{QMZ}
Z(\lambda, \chi) = \la \psi| e^{i\lambda {\hat A}} e^{i\chi {\hat B}} | \psi \ra.
\ee
Note the exponentials are factorized to respect the potential non-commutation of $\hat A$ and $\hat B$ as well as the chosen operator ordering.

We can build on the analysis of the previous section to extract the pseudo-distribution from experimentally measured correlation functions. We define two real-valued $d \times d$ Vandermonde matrices 
\begin{eqnarray}
    {\bf V} = & \begin{pmatrix} 1 & 1 & \ldots  & 1 \\
a_1 & a_2 & \ldots & a_{d} \\
a_1^2 & a_2^2 & \ldots  & a_{d}^2 \\
\vdots & \vdots &   & \vdots \\
a_1^{d-1} & a_2^{d-1} & \ldots & a_{d}^{d-1}
\end{pmatrix}, \\
{\bf W} = & \begin{pmatrix} 1 & 1 & \ldots  & 1 \\
b_1 & b_2 & \ldots & b_{d} \\
b_1^2 & b_2^2 & \ldots  & b_{d}^2 \\
\vdots & \vdots &   & \vdots \\
b_1^{d-1} & b_2^{d-1} & \ldots & b_{d}^{d-1}
\end{pmatrix}.
\end{eqnarray}
We now let the indices $n, m$ run from 0 to $d-1$.  Viewing the collection of $d\times d$ correlation functions (\ref{cnmsum}) as a matrix ${\bf C}$, and the pseudo-distribution $Q_{i,j}(\psi)$ as a matrix ${\bf Q}$,
we have the equation
\be
{\bf C} = {\bf V} \cdot {\bf Q} \cdot {\bf W}^T.
\ee
Thus, the pseudo-distribution matrix ${\bf Q}$ can be determined empirically from the matrix of correlation functions together with the (known) Vandermonde matrices' inverses:
\be
{\bf Q} = {\bf V}^{-1} \cdot {\bf C} \cdot ({\bf W}^{-1})^T.
\ee
This completes the empirical construction of $\bf{Q}$.

The characteristic function in Eq.~\eqref{QMZ} takes the form of an expectation value of two exponentials. We can impose the requirement that the underlying pseudo-distribution should reproduce this expectation value
\be
Z(\lambda, \chi) = \sum_{i, j} Q_{i,j}(\psi) e^{i \lambda a_i + i \chi b_j} . \label{pdz}
\ee
Here, we have used the property that the eigenvalues of a function of an operator are equal to the same function of the eigenvalues of the original operator. 
  From a quantum mechanical point of view, we can insert $ {\bf 1}  = \sum_i \ket{a_i}\bra{a_i}$ and $ {\bf 1}  = \sum_j \ket{b_j}\bra{b_j}$ into Eq. \eqref{QMZ}:
\be
Z(\lambda, \chi) = \sum_{i, j} \la \psi | a_i \ra \la a_i | b_j \ra \la b_j | \psi \ra
e^{i \lambda a_i + i \chi b_j} .  \label{qmz}
\ee
Comparing  Eqs.~(\ref{pdz}) and (\ref{qmz}), we identify the underlying pseudo-distribution as the (complex-conjugated) standard Kirkwood-Dirac pseudo-distribution:
\be
Q_{i,j}(\psi) = \la \psi | a_j \ra \la a_j | b_i\ra \la b_i | \psi\ra = K^*_{i,j}(\psi).  \label{standkd}
\ee
Thus, we have established a correspondence between the Kirkwood-Dirac distribution and an empirically measured one from a set of correlation function with minimal theoretical input.

\subsection{Multiple-observable measurement}  \label{multiple}
The previous section generalizes in a straightforward way to an $N$-point correlation function of the form
\be
\label{cgen}
C_{m_1,m_2, \ldots m_N} = \la \psi| {\hat O_1}^{m_1} {\hat O_2}^{m_2} \ldots {\hat O_N}^{m_N}  | \psi\ra.
\ee
We introduce a general characteristic function similar to the previous section
\be
\label{Qgen}
Z(\lambda_1, \lambda_2, \ldots, \lambda_N ) = \la \psi| e^{i\lambda_1 {\hat O}_1} e^{i\lambda_2 {\hat O}_2}
\cdots e^{i\lambda_N {\hat O}_N}
| \psi \ra.
\ee
Inserting the assumed pseudo-distribution yields for the quantum characteristic function
\begin{eqnarray}
&Z&(\lambda_1, \lambda_2, \ldots, \lambda_N )  \label{pdz} \\ &=& \sum_{i_1, i_2, \ldots i_N} Q_{i_1,i_2, \ldots, i_N}(\psi) e^{i \lambda_1 o_{i_1} + i \lambda_2 o_{i_2} + \ldots + i \lambda_N o_{i_N}} . \nonumber
\end{eqnarray}
Similar arguments as in the previous section lead to the generalized KD pseudo-distribution, $Q=K$, where
\be
K_{i_1,i_2, \ldots i_N}(\psi) = \la \psi | o_{i_1} \ra \la o_{i_1} |  o_{i_2} \ra \la o_{i_2}  | \cdots  | o_{i_N} \ra \la o_{i_N}| \psi\ra,  \label{standkd}
\ee
where the eigenvalues and eigenstates are given by ${\hat O} | o_i\ra = o_i |o_i\ra$ for each operator considered.

\section{Continuous variables}
\label{continuous}

We now generalize this analysis to continuous-variable systems represented on infinite-dimensional Hilbert spaces.  Now, the sums become integrals, and the matrices become infinite-dimensional. 
We let the eigenvalues and eigenvectors of a continuous operator $\hat {\cal O}$ be represented by $\lambda$ and $\ket{\lambda}$, respectively, such that $\hat {\cal O} = \int \lambda \ket{\lambda} \bra{\lambda } d\lambda$. 

In analogy to the previous sections, we consider measurements of the conditional expectation value of $\la {\cal \hat O}^n \ra^{\psi}_{\phi}$ of the $n$th power of the operator $\hat {\cal O}$. Here, $\ket{\psi}$ and $\ket{\phi}$ represent the pre-selected and post-selected states, respectively. For a classical continuous-variable system, one has that $\la {\cal \hat O}^n \ra^{\psi}_{\phi} = \int \lambda^n p(\lambda|\psi, \phi)$, where $p(\lambda|\psi,\phi)$ is some conditional probability density. As before, we investigate whether the structure of the classical formula can be preserved in the quantum scenario. We assume the existence of a continuous-variable pseudo-distribution $q(\lambda|\psi,\phi)$ such that 
\be
\la {\cal \hat O}^n \ra^{\psi}_{\phi} = \int d\lambda \lambda^n q(\lambda|\psi, \phi). \label{o^n}
\ee
 
In the discrete-variable analysis, we saw that we could specify the pseudo-distribution if we measured the conditional expectation values of $d-1$ powers of the operator of interest. When $d$ is infinite, this is, naturally, infeasible. Instead, we consider the weak characteristic function, from which the conditional expectation values can be derived:
\be
Z_{\psi,\phi}(\chi) = \sum_{n=0}^\infty \frac{i^n}{n!}  \chi^n \la {\cal \hat O}^n \ra^{\psi}_{\phi}
= \int d\lambda e^{i \chi \lambda} q(\lambda|\psi,\phi).
\label{weakgf}
\ee
Here, we used  Eq.~(\ref{o^n}) and summed over the scalar series as the pseudo-distribution factors out of the sum.

If we had experimental data of $Z_{\psi,\phi}(\chi)$, then we could determine the underlying pseudo-distribution, $q(\lambda|\psi,\phi)$, via an inverse Fourier transform. From the form of Eq. \eqref{weakgf} it appears, however, that to find $Z_{\psi,\phi}(\chi)$ one must either know $q(\lambda|\psi,\phi)$ \textit{a priori} or measure the infinite number of conditional expectation values, $\la {\cal \hat O}^n \ra^{\psi}_{\phi}$. This turns out to be false. From quantum mechanics, we know that the weak value \cite{aharonov1988result,dressel2014colloquium} is given by $\la {\cal \hat O}\ra^{\psi}_{\phi} = \la \phi | {\cal \hat O} | \psi \ra / \la \phi | \psi \ra $. Substituting this expression into the left equality of Eq. \eqref{weakgf} yields  
\be
Z_{\psi,\phi}(\chi) =  \left\la \exp \left(i \chi {\cal \hat O}\right)\right\ra^{\psi}_{\phi} . \label{op}
\ee
This last result suggests a deep connection to quantum modular variables (see \cite{Aharonov1969modular, lobo2014modular}), but is also of direct experimental significance. As we shall demonstrate in the following section, Eq.~\eqref{op} enables the measurement of the pseudo-distribution's characteristic function via a {\it single} experimental setup with a tunable parameter $\chi$. In other words, one does not have to experimentally measure an infinite number of moments to specify the characteristic $Z_{\psi,\phi}(\chi)$. 
Instead, a well-defined operational procedure gives a dramatic simplification in the empirical construction of the pseudo-distribution, which indeed can be determined by taking the inverse Fourier transform of $Z_{\psi,\phi}(\chi)$ using the representation (\ref{op}). 


Next, we theoretically determine the form of the pseudo-distribution that would be determined by the above-outlined procedure. Consider inserting the continuous resolution of identity ${\bf 1} = \int d\lambda |\lambda \ra \la \lambda |$ into the expectation value in Eq. \eqref{op}. This yields
\be
\label{Zweak}
Z_{\psi,\phi}= \int d\lambda e^{i \chi \lambda} \frac{\la \phi | \lambda \ra \la \lambda | \psi \ra}{\la \phi | \psi \ra}.
\ee
 Comparing  Eqs. \eqref{weakgf} and \eqref{Zweak}, we can identify the unique expression for the pseudo-distribution:
 \be
 q(\lambda |\psi, \phi) = \frac{\la \phi | \lambda \ra \la \lambda | \psi \ra}{\la \phi | \psi \ra}.
 \ee
 We can interpret this pseudo-distribution as the weak-valued projector corresponding to the eigenvalue $\lambda$.  Moreover, this expression is precisely the conditional continuous-variable Kirkwood-Dirac pseudo-distribution (see Eq.~(25) of Ref.~\cite{arvidsson2024properties}).

  \begin{figure}
  \begin{center}
\includegraphics[width=0.5\textwidth]{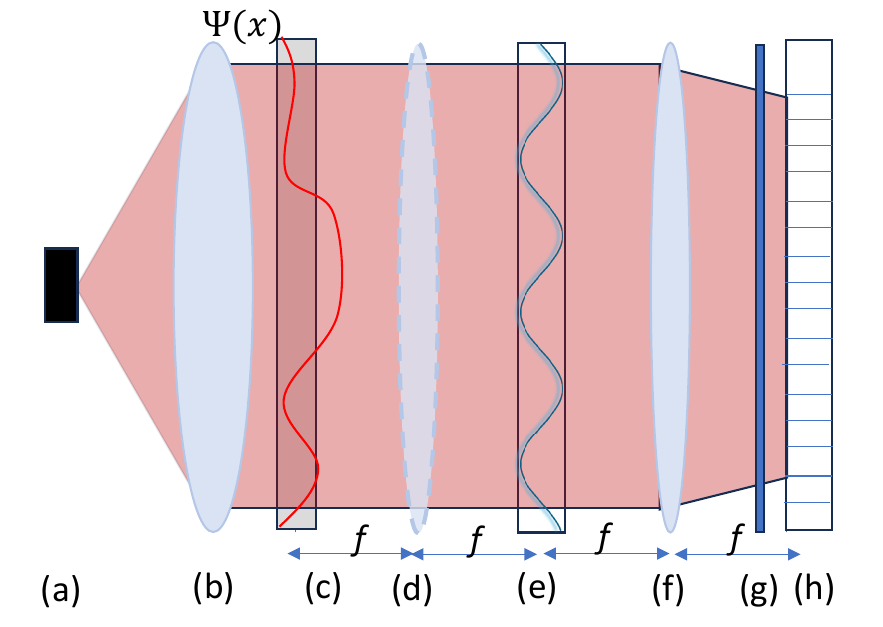}
  \end{center}
  \caption{Schematic of the proposed experiment to directly measure the pseudo-distribution characteristic function.  The optical configuration consists of these elements: (a) single-photon source and (b) beam expanding lens, (c) mask to imprint the quantum state of the single-photon $\Psi(x)$, (d) optional focusing lens with focal length $f$, (e) weak periodically modulated birefringence, (f) focusing lens with focal length $f$, (g) polarizer, (h) Single photon camera.  The camera is placed in the Fourier plane of the lens.}
    \label{fig:exp-setup}
  \end{figure} 

\section{Proposed experiment for direct measurement of the characteristic function}
\label{experiment}
The experimental schematic we propose is shown in Fig.~\ref{fig:exp-setup}.  There, we choose the operator ${\cal \hat O}$ to be $ {\hat x}$, the position operator, and the final measurement basis $\ket{b_j}$ to be the transverse momentum states $|p\ra$. After imprinting the desired quantum state (indicated as a red curve) on a single photon in steps (a-c), the position characteristic function can be weakly measured by slightly rotating the polarization of the single photon continuously as a function of position, step (e) as a sinusoid.  This can be accomplished (for example) using a parallel aligned nematic liquid crystal spatial light modulator (SLM).
  The spatial frequency of the polarization pattern can be continuously controlled through the spatial light modulator \cite{moreno2012complete}; alternatively, an ancillary {\em spatial} direction could be use to multiplex even this step.  The final measurement of momentum is accomplished via a Fourier transform lens (f).   The final polarizer (g) and single-photon camera (h) complete the weak measurement of the characteristic function, $\sin(k {\hat x} + \phi)$, where $k$ is the SLM frequency.  Measurements with different polarization settings are needed to reconstruct both real and imaginary parts of the conditioned average (and can also be performed in parallel).  The lens (d) is inserted if one wish to measure momentum instead, as discussed in the next section. 
  Data of the single-photon histogram is taken for a range of different values of the SLM frequency for both the sine and cosine quadratures.  The pseudo-distribution can then be produced by numerical inverse Fourier transform with different values of the final momentum.  We note that an alternative method for directly measuring the KD pseudo-distribution was proposed by scanning a sliver of a half-wave plate through the beam-profile, which weakly measures a position projector \cite{lundeen2012procedure}.  Our method is conceptually different and more efficient than the direct position projection measurement in that it measures a global property of the state ``all at once''.  Furthermore, the period of the SLM can be modulated spatially or temporally, enabling extraction of the weak characteristic function to be parallelized --- in contrast to the raster scan of a sliver of birefringent crystal to directly make weak measurements of the 
  position projectors everywhere. 

\subsection{First x, Then p}
This procedure (without lens (d)) is equivalent to projecting on a final momentum $| p \ra$. We can realize this measurement by transforming our state to the Fourier plane via a lens with focal length $f$. Then, the final position $x_f$ corresponds to the momentum as $x_f = \lambda f p$, where $\lambda$ is the wavelength of the light.  Thus, the detector measures the conditioned average
\be
Z_{p, \psi} = \frac{\la p | e^{i k {\hat x}} |\psi\ra }{\la p | \psi\ra} = \frac{\la p - k |\psi\ra }{\la p | \psi\ra}.
\ee
In the second equality, we used the fact that the position operator is the generator of translations of the momentum.  We recognize the characteristic function as ratios of momentum space wavefunctions at different arguments.  According to the prescription above, the pseudo-distribution at position $x$ is given by the inverse Fourier transform,
\be
Q(x | \psi, p) = {\cal F}^{-1}_{k \rightarrow x} Z_{p, \psi}.
\ee
Consistency can be checked by taking the inverse Fourier transform of the characteristic function directly to find
\begin{eqnarray}
Q(x | \psi, p) &=& \frac{\la p | \delta(x-{\hat x}) |\psi\ra }{\la p | \psi\ra} \label{stepa},\\
&=& \int dy \frac{\la p | \delta(x-{y}) |y\ra \la y| \psi\ra }{\la p | \psi\ra} \label{stepb},\\
&=& \frac{\la p |x\ra \la x| \psi\ra }{\la p | \psi\ra}.\label{stepc}
\end{eqnarray}
In step (\ref{stepa}) we used the fact that the inverse Fourier transform of the complex exponential is the delta-function; in step (\ref{stepb}), we inserted a complete set of position states $y$, and applied the position operator to ket $|y\ra$; and in step (\ref{stepc}), we performed the $y$ integral.  As expected, we find that the pseudo-distribution for position $x$ is the weak-value of the position projector at $x$.

It is also possible to extract the joint pseudo-distribution by considering the joint characteristic function
\be
Z(\lambda, k) = \la \psi | e^{i \lambda {\hat p}} e^{i k {\hat x}} | \psi\ra = \int dx dp\, Q(x, p | \psi) e^{i \lambda p + i k x}.
\ee
Inserting a complete set of position and momentum states, we can give the quantum mechanical prediction
\be
Z(\lambda, k) = \int dx dp \la \psi | p \ra \la p | x\ra \la x | \psi\ra  e^{i \lambda p + i k x}.
\ee
Thus, we identify the pseudo-distribution to be
\be
K(x, p|\psi) = \la p | x\ra \la x|\psi\ra \la \psi|p\ra, \label{qkd}
\ee
which is exactly the standard Kirkwood-Dirac pseudo-distribution \cite{arvidsson2024properties}.  This distribution can be constructed from the above experiment, considering also the probability of measuring the final momentum (i.e. not conditioning on the final momentum), weighting the conditional pseudo-distribution (\ref{stepc}) by the probability of the final momentum $p$, that is, $\la \psi | p\ra \la p | \psi\ra$.

\subsection{First p, Then x}
Alternatively, lens (d) can be inserted into the experimental apparatus in Fig.~\ref{fig:exp-setup}.  In this case, the two later lenses form a 4$f$ imaging system:  the first lens takes the waveform to the Fourier plane, so the periodic modulation is now applied in momentum space, corresponding to a weak measurement of the characteristic function $\sin( \lambda {\hat p}+\phi)$.  The second lens returns the waveform back to position space, so the final measurement is on the pixel position $x$.   The generating function of the conditional moments of momentum is then given by
\be
{\tilde Z}_{x, \psi} = \frac{\la x | e^{i \lambda {\hat p}} |\psi\ra}{\la x | \psi\ra}.
\ee
In a similar way as in the previous subsection, we can also find the joint characteristic function to be
\be
{\tilde Z}(k, \lambda) = \la \psi | e^{i k {\hat x}} e^{i \lambda {\hat p}} | \psi\ra = \int dx dp\, {\tilde Q}(x, p | \psi) e^{i \lambda p + i k x},
\ee
which reverses the ordering of position and momentum.
Here, we introduced the reversed order pseudo-distribution ${\tilde Q}(x,p)$, for which the quantum mechanical prediction is 
\be
{\tilde K}(x, p | \psi) = \la x| p\ra \la p | \psi\ra \la \psi | x\ra. \label{qtildekd}
\ee

\subsection{Direct Measurement of the Canonical Commutation Relation}
\label{ccr}
We note that the quantum mechanical expressions satisfy ${\tilde K}(x, p | \psi) =  K(x, p | \psi)^\ast$.  If we take the difference of the correlation function of $x p$ in the first and the second experiment, we have the expression
\be
\la x p \ra_{\tilde K} - \la x p \ra_{K} = \int dx dp \ x p \ 2i {\rm Im} [{ \tilde K}(x, p | \psi)] \label{diff}.
\ee
This expression can be simplified by writing the KD distributions in terms of the momentum and position projection operators ${\hat \pi}_p = |p\ra \la p| , {\hat \pi}_x = | x\ra \la x|$.  Expressions (\ref{qkd}, \ref{qtildekd}) imply that
\begin{eqnarray}
 \la x p \ra_K = \int dx dp \ xp \ \la \psi | {\hat \pi}_p {\hat \pi}_x | \psi \ra =  \la \psi | {\hat p} {\hat x} | \psi \ra,  \\
 \la x p \ra_{\tilde K} = \int dx dp \ xp \ \la \psi |  {\hat \pi}_x {\hat \pi}_p | \psi \ra =  \la \psi | {\hat x} {\hat p} | \psi \ra.
\end{eqnarray}
Thus, the difference (\ref{diff}) is given by
\be
\la x p \ra_{\tilde K} - \la x p \ra_{K} = \la \psi | {\hat x} {\hat p} - {\hat p} {\hat x} | \psi\ra = i \hbar,\label{ccr}
\ee
where the last step follows from the canonical commutation relation, and must be true for every quantum state.  This is a powerful and universal prediction from quantum physics that can be directly probed, constituting a direct measurement of the canonical commutation relation (cf. \cite{zavatta2009commutation}).  The predicted equivalence of (\ref{ccr}) and (\ref{diff}) also indicates that KD distribution functions must take on complex values to satisfy the canonical commutation relation.



\section{Conclusions}
We have developed a theory of quantum pseudo-distributions generated via measurements of successive moments of quantum operators, or equivalently, in terms of its characteristic function.  By enforcing analogies to classical statistics, we provide a theory-agnostic way to construct pseudo-distribution representations of quantum experiments.  In the discrete case, we find that the inverses of Vandermonde matrices play a key role in calculating the inferred pseudo-distribution, while in the continuous case, the pseudo-distribution can be found by the inverse Fourier transform of the characteristic-function data.  A proposed experiment to demonstrate this method was presented, which is related to ``direct'' measurements of the wavefunction \cite{lundeen2011direct}.  A key difference is the ``all at once'' nature of the measurement encodes global properties of the quantum state.  We also proposed a method for directly verifying the canonical commutation relation via reversed measurements of momentum and position.  Our theory shows that if one requires the classical description of these experiments to carry over to the quantum regime, one is forced to update the classical joint probability distribution to the Kirkwood-Dirac pseudo-distribution.

As an outlook for future work, we note that our approach may provide a more efficient way of doing quantum state tomography for continuous variable systems. This is because the pseudo-distribution gives an equivalent encoding of information from the quantum state, see Eq.~(\ref{reconstruct}).  The method described may be more efficient from a resource perspective than other common methods such as the inverse Radon transformation or maximum-likelihood techniques (where the reconstruction of non-gaussian states can be exponentially demanding \cite{lvovsky2009continuous, rozema2014CV, mele2025CV}).

We note that a recent article \cite{spriet2025specialkirkwooddiracdistributions} also touches on criteria for uniquely identifying the KD distribution as the natural pseudo-distribution. The authors show that the KD distribution is the only Born-rule compatible pseudo-distribution for which the associated conditional expectation coincides with the best predictor, as it does classically.

Note added: During the preparation of this manuscript, we learned that independent work on a related topic has been done by Haruki Emori.

\section{Acknowledgments}
This work was supported by NSERC Discovery under Grant No. RGPIN-2020-05767 and by the John Templeton Foundation under grant ID 63209.  ANJ acknowledges support from the Kennedy chair in physics at Chapman University.  We thank Jeff Lundeen, Tryphon Georgiou, Ralph Sabbagh, Justin Dressel and Haruki Emori for helpful discussions and valuable comments on the manuscript.

\end{document}